\pdfoutput=1
\documentclass[twocolumn, tighten]{aastex61}
\usepackage{amsmath}

\shorttitle{Maximized data rate for lightweight space-probes}
\shortauthors{Hippke}

\begin{document}
\title{Interstellar communication. I. Maximized data rate for lightweight space-probes}

\author[0000-0002-0794-6339]{Michael Hippke}
\affiliation{Sonneberg Observatory, Sternwartestr. 32, 96515 Sonneberg, Germany}
\email{hippke@ifda.eu}

\begin{abstract}
Recent technological advances could make interstellar travel possible, using ultra-lightweight sails pushed by lasers or solar photon pressure, at speeds of a few percent the speed of light. Obtaining remote observational data from such probes is not trivial because of their minimal instrumentation (gram scale) and large distances (pc). We derive the optimal communication scheme to maximize the data rate between a remote probe and home-base. he framework requires coronagraphic suppression of the stellar background at the level of $10^{-9}$ within a few tenths of an arcsecond of the bright star. Our work includes models for the loss of photons from diffraction, technological limitations, interstellar extinction, and atmospheric transmission. Major noise sources are atmospheric, zodiacal, stellar and instrumental. We examine the maximum capacity using the ``Holevo bound'' which gives an upper limit to the amount of information (bits) that can be encoded through a quantum state (photons), which is a few bits per photon for optimistic signal and noise levels. This allows for data rates of order bits per second per Watt from a transmitter of size 1\,m at a distance of $\alpha$\,Centauri (1.3\,pc) to an earth-based large receiving telescope (E-ELT, 39\,m). The optimal wavelength for this distance is 300\,nm (space-based receiver) to 400\,nm (earth-based) and increases with distance, due to extinction, to a maximum of $\approx3$\,$\mu$m to the center of the galaxy at 8\,kpc.
\end{abstract}

\section{Introduction}
Interstellar travel became technologically plausible in the 1950s, when the energy release of thermonuclear fusion was observed in the first hydrogen bombs. First studies were based on the idea of a pulse drive, directly propelled by the explosions of atomic bombs behind the craft \citep{1965Sci...149..141D, 1968PhT....21j..41D}, evolving into a direct fusion rocket \citep{1978JBIS...31S...5B}. These designs were manned interstellar arks with masses of order 10 million tonnes and speeds of 10\,\% the speed of light.

Classical rockets, both chemical and nuclear, suffer from the limitations imposed by Tsiolkovsky's rocket equation \citep{1992CeMDA..53..227P}: if a rocket carries its own fuel, the ratio of total rocket mass versus final velocity is an exponential function, making high speeds extremely expensive. A different method, which does not require the fuel to be accelerated with the ship, has been proposed by Johannes \citet{kepler1604}. After observing a comet, he suggested that the cometary tail points away from the sun due to a ``breeze'', and proposed to ``provide ships or sails adapted to the heavenly breezes, and there will be some who will brave even that void.''. James Clerk Maxwell predicted that radiation carries momentum and exerts pressure: ``Hence in a Medium in which waves are propagated there is a pressure in the direction normal to the wave, and numerically equal to the energy in unit of volume'' \citep{Maxwell1873,maxwell1990scientific}.

\citet{1967Natur.213..588R} noted that there was no obvious way to decelerate the spacecraft at the target star system. Only recently, \citet{2017ApJ...835L..32H} and \citet{2017arXiv170403871H} suggested to decelerate using the stellar radiation and gravitation in a maneuver they referred to as photogravitational assist. A project by the ``Breakthrough Initiatives''\footnote{\url{http://breakthroughinitiatives.org}} provides monetary support (of order 100\,m USD) for research on gram-scale robotic spacecrafts, using a light sail for propulsion \citep{2016arXiv160401356L,2017Natur.542...20P}.

Between ``Project Orion'', and the ``nanocraft concept'', there is a factor of $10^{13}$ in weight. The smaller weight results in lower build- and launch costs, a benefit that could make such a mission affordable within the current century. When we compare the early studies with the most recent concept, we have to distinguish that the main purpose of interstellar travel shifted from colonization of exoplanets with human (biological) settlers to unmanned research probes, taking spectroscopic and photographic measurements of the putative biological environment on potentially habitable exoplanets. Software and hardware engineering has made sufficient progress since the 1960s that such probes can be highly autonomous. Consequently, the required mass for probes can be reduced.

Our benefit from autonomous interstellar probes is purely in the information they send back to us. Thus we shall seek to maximize the amount of information we can obtain from them. A major issue is that these probes are designed to be very light-weight, and thus limited in terms of power. While traditional, fusion-based concepts proposed the use of high \citep[MW,][]{2016JBIS...69...278G} power at GHz frequencies for data transfer, small sailing probes can not have a fusion reactor on board and will have to rely on photovoltaic energy, which delivers of order kW per square meter surface area. In the current era of high resolution video, a high data rate to transfer spectacular observations of an alien world could be important for the public reception of such a mission, and thus its financial funding. It is therefore crucial to optimize interstellar communication, precisely the data rate, to maximize the data volume of scientific and public data.

In this work, paper I of the series, our contributions are: (1) to introduce the variables in the framework of data transfer between telescopes; (2) to assess limiting factors such as extinction, noise, and technological constraints; and (3) to calculate optimal frequencies and achievable data rates for exemplary cases.

\section{Method to calculate data rates}
The free-space photon flux $F$ received from a telescope at distance $d$ can be calculated as \citep{kaushal2017free}:

\begin{equation}
\label{eq1}
F = \frac{P_{\rm t}}{\pi h f (\theta d)^2}
\end{equation}

where $P_{\rm t}$ is the transmitted power, $f$ the photon frequency, and $h$ Planck's constant ($\approx6.626\times10^{-34}$\,J\,s). The (half) opening angle of the diverging light beam is $\theta_{\rm d} = Q_{\rm R} \lambda/D_{\rm t}$ (in radians) with $Q_{\rm R}\approx1.22$ for a diffraction limited circular transmitting telescope of diameter $D_{\rm t}$ \citep{rayleigh}, and $\lambda=c/f$ with $c$ as the speed of light in vacuum ($299,792,458$\,m\,s$^{-1}$). In a receiving telescope with aperture $A_{\rm R}=\pi D_{\rm r}^2/4$ we obtain the flux

\begin{align}
\label{eq2}
F_{\rm r} = \frac{P_{\rm t}}{\pi h f (Q_{\rm d} \lambda / D_{\rm t})^2 d^2} \times \frac{\pi D_{\rm r}^2}{4} \nonumber \\
= \frac{P_{\rm t} D_{\rm t}^2 D_{\rm r}^2}{4 h f Q_{\rm d}^2 \lambda^2 d^2} &({\rm s}^{-1}).
\end{align}

This assumes a uniform plane-wave illumination. A telescope with central obscuration and plane-wave gaussian-beam illumination has been calculated by \citet{1974ApOpt..13.2134K}, and the flux loss from pointing errors by \citet{1987ApOpt..26.2055M}; but these secondary effects will be neglected here. For a laserbeam, the narrower ``waist'' leads to an intensity pattern with a characteristic angular beam size given by \citep{duarte2015tunable,2015PASP..127..540T} $\theta_{\rm L} = Q_{\rm L} (2/\pi) \lambda/D$, or $\theta_{\rm L} / \theta_{\rm d} \approx 0.5$, which leads to a tightening of the beam. Note that a laserbeam shape is not maintained in systems where laser light is broadened with a beam expander and then focused with a telescope, and so we neglect this possibility here.

The widely used approximation\footnote{Approximations and mistakes in the literature will be discussed in section~\ref{lit}.} of the diffraction-limited aperture, $\theta \approx \lambda / D$, leads to an overestimate of the received photon flux on the receiver side by $\approx49$\%. This can be verified by setting $Q_{\rm R}=1$ versus $Q_{\rm R}=1.22$ numerically. The considerable difference comes from the fact that $\theta$ enters the equation through the inverse square law. The precise value, $\theta = 1.2196..\lambda / D$ comes from the Fraunhofer diffraction where this number is the first zero of the order-one Bessel function of the first kind, $J_1(x)/\pi$.

Several factors will constrain the achievable data rates. Regarding the loss of photons, the most important are interstellar extinction (section~\ref{loss_ext}), of which we denote the surviving fraction as $0<S_{\rm E}<1$. For ground-based telescopes, atmospheric transmission allows for the reception of another fraction of photons, $0<S_{\rm A}<1$ (section~\ref{atmo}). The receiver efficiency is denoted as $0<\eta<1$. Technological constraints on the telescopes will be denoted as $Q$ (section~\ref{sec_tech_limit}). Other small factors, such as scintillation and scattering (section~\ref{scatter}), might play a role and can be included in calculations in a similar manner, but we neglect them here for brevity. The major noise sources are atmospheric sky background (section~\ref{sky}), zodiacal light (section~\ref{noise_zodi}), and others (sections~\ref{star},~\ref{instrumental_noise}).

\subsection{Channel capacity for a coherent wave}
\label{channel_capacity}
We now define the theoretical maximum data rates based on frequency, signal and noise. For completeness, we will first discuss the optimum case where the number of photons received is sufficiently large to form a coherent wave. While this might not be realistic for most schemes of interstellar communication (section~\ref{results}), it is useful to define the classical upper bound. The maximum rate at which information can be transmitted over a communications channel is \citep{Shannon1949}:

\begin{equation}
\label{shannon}
C=B \log_2 \left(1+\frac{S}{N}\right)
\end{equation}

where $C$ is the channel capacity (in bits per second), $B$ is the bandwidth of the channel (in Hertz), $S$ is the average signal power (in Watt) and $N$ is the average gaussian noise (in Watt). The bandwidth is the difference between the highest ($f_{\rm H}$) and lowest ($f_{\rm L}$) frequency in a continuous set of frequencies.

To compare data rates for different frequencies, we can approximate bandwidth with frequency by taking a constant fractional bandwidth, $b$. With $f_{\rm C}$ as the center frequency, we can define $b=(f_{\rm H}-f_{\rm L})/f_{\rm C}$. With a value of e.g. $b=0.1$, we can approximate $B\approx c/\lambda$ (in Hz).

Channel capacity is proportional to frequency and to the logarithm of S/N. These relations suggest that the frequency should be increased to the practical maximum, and that the signal power should merely be increased to overcome noise, with little benefit beyond.

If the received number of photons (after extinction and other losses) is sufficiently large to form a coherent wave, we can plug Eq.~\ref{eq2} (as the signal $S$ in photons per second) into \ref{shannon}, and define the noise equally in photons per second:

\begin{equation}
\text{DSR}\textsubscript{c} = \frac{c}{\lambda} \log_2 \left(1+\frac{F_{\rm r}}{N_{{\rm \gamma}}}\right)
\end{equation}

where $N_{\rm \gamma}$ is the number of photons ($\gamma$) from noise per second (physical and instrumental). Then, the data signaling rate for the case of a continuous wave, $\text{DSR}\textsubscript{c}$, can be conveniently calculated in units of bits/s if $P$ is in Watt.

Intuitively, one would assume that at least one photon is required to transfer one bit of information, but this is incorrect: More than one bit can be encoded per photon. This is done with a modulation scheme to define an alphabet, often using a combination of polarization, phase, frequency and amplitude modulation \citep[e.g.,][]{1995ASPC...74..369J}. Each symbol of such an alphabet can encode several bits, scaling with the logarithm to base 2 of the number of members. This is called spectral efficiency and is measured in (bits/s)/Hz. Modulation rate, spectral efficiency and data rate can be increased for a constant bandwidth at the cost of an exponential rise in SNR, or, for a constant noise level, in an exponential increase in $P$.

For the extreme case of communication with negligible losses (e.g., $d \to 0$), Eq.~\ref{shannon} suggests the use of infinitely high bandwidth. However, infinite frequencies (and infinite capacity) are unphysical. In the classical sense, the limit comes from the fact that an increase in bandwidth also increases noise power (Shannon's power efficiency limit). A noiseless channel has infinite capacity: with Eq.~\ref{shannon} we have $C=B \, \log_2 \, (1+\infty)=\infty$. However, in reality noise is never zero because photons are quantized (section~\ref{instrumental_noise}). Then, the capacity of Shannon's limit becomes \citep[][p. 5-117]{chitodecommunication}:

\begin{equation}
\lim_{B \to \infty} C = \frac{S}{N} \log_e 2 \approx 1.44 \frac{S}{N}
\end{equation}

In the framework of quantum state propagation, any transmission system can exchange only a limited (quantized) amount of information in a given time frame \citep{1978ITIT...24..657Y}, and is thus limited by physical resources \citep{1981PhRvD..23..287B}. Therefore, increasing frequency to infinity does not increase data rate to infinity \citep{2004PhRvL..92b7902G}.

\begin{figure}
\includegraphics[width=\linewidth]{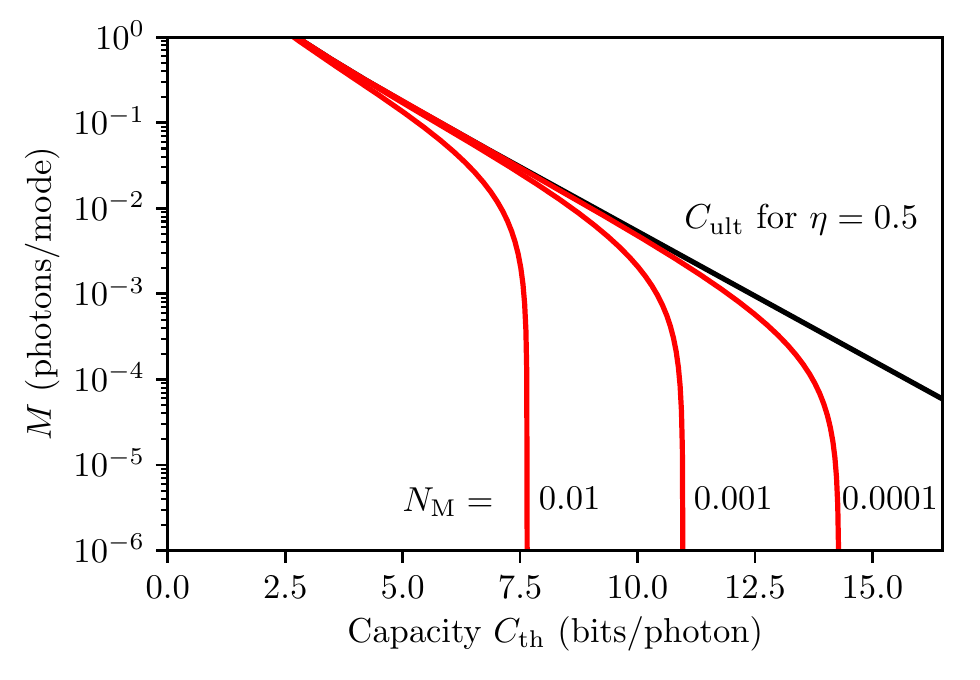}
\caption{\label{figure_holevo}Capacity $C_{\rm th}$ in bits per photon as a function of the number of photons per mode, $M$. The larger the number of modes, the more bits can be encoded per photon, however the ultimate bound (black) is logarithmic. When accounting for thermal noise per mode $N_{\rm M}$ (fractions in the plot), the limits are even tighter (red lines).}
\end{figure}

\subsection{The photon limited case}
The limit for Eq.~\ref{shannon} only applies if the number of photons is sufficiently large to form a coherent wave. In many schemes for interstellar communication (section~\ref{results}), the data rate is photon-limited. Then, Holevo's bound \citep{holevo1973bounds} establishes the upper limit to the amount of information which can be transmitted with a quantum state. It applies independently from the frequency of the wave, and assumes that a number (quantity) of modes can be used per photon, which originate from the photons' dimensions, namely polarization, frequency and time of arrival. The inverse of this quantity, $M$, is the number of photons per mode. Then, as shown by \citet{2004PhRvL..92b7902G}, the ultimate quantum limit of bits per photon can be expressed as:

\begin{equation}
C_{\rm ult}=g(\eta M)
\end{equation}

where $\eta$ is the receiver efficiency and $g(x)=(1+x) \log_2 (1+x)-x \log_2 x$ so that $g(x)$ is a function\footnote{An introduction into quantum information theory and the usual notation can be found in \citet{2014PhRvA..89d2309T}.} of $\eta \times M$. In the presence of thermal noise, it was conjectured \citep{2004PhRvA..70c2315G} and recently proven \citep{2014NaPho...8..796G} that the capacity is:

\begin{equation}
\label{thermal_holevo}
C_{\rm th}=g(\eta M + (1-\eta) N_{\rm M}) - g((1-\eta)N_{\rm M})
\end{equation}

where $N_{\rm M}$ is the average number of noise photons per mode. It is an open question if the maximum can fully, or only approximately be achieved in practice \citep{2012arXiv1202.0518W,2012arXiv1202.0533G}. The achievable capacity is shown for a wide range of modes in Figure~\ref{figure_holevo}. It is clear that even large numbers of modes and small fractional noise increase the number of bits per photon only within a factor of a few.

We can multiply Eq.~\ref{eq2} and Eq.~\ref{thermal_holevo} to calculate the data rate for the photon limited case of two communicating telescopes:

\begin{equation}
\label{bits_holevo}
\text{DSR}\textsubscript{$\gamma$} = C_{\rm th} F_{\rm r}
\end{equation}

where \text{DSR}\textsubscript{$\gamma$} is in units of bits/s when $P$ is in Watt. It assumes that the free path loss caused by $\eta, d, S_{\rm E}, S_{\rm A}$ is known and accounted for in the encoding scheme. Variations and uncertainties in the number of received photons can be treated as an additional noise source, but optimal encoding schemes will be neglected in this paper. In the following sections, we will discuss the values in these equations.

\begin{figure}
\includegraphics[width=\linewidth]{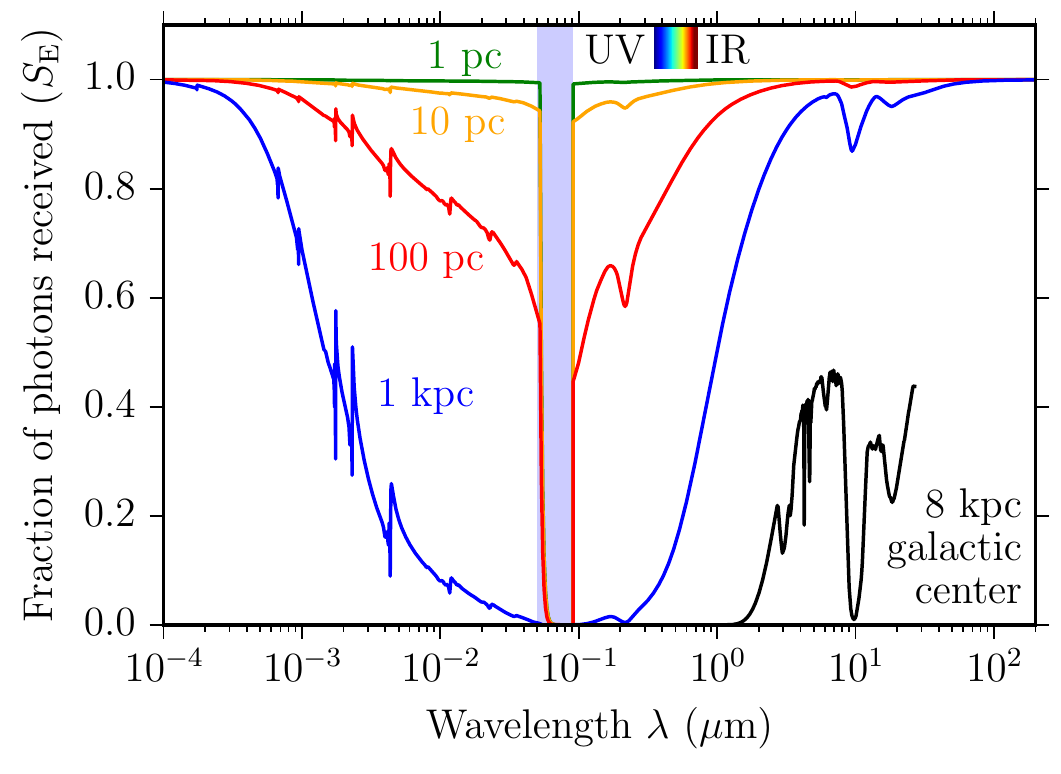}
\caption{\label{extinction}Fraction of photons that defies interstellar extinction ($S_{\rm E}$), as a function of wavelength $\lambda$, shown for different distances. The shaded area represents the Lyman continuum ($\approx50-91.2$\,nm) which is opaque even for the closest stars due to the ionization of neutral hydrogen \citep{1959PASP...71..324A,2000ApJ...542..914W,2007eua..book.....B}.}
\end{figure}

\section{Signal losses}
\subsection{Loss of photons from extinction}
\label{loss_ext}
From the IR to the UV, extinction is caused by the scattering of radiation by dust, while at wavelengths shorter than the Lyman limit (91.2\,nm), extinction is dominated by photo-ionisation of atoms \citep{1996Ap&SS.236..285R}. For short interstellar distances, extinction in the optical is small, $\approx0.1$\,mag within 100\,pc, $0.05-0.15$\,mag out to 200\,pc \citep{1998A&A...340..543V}. It is much larger towards the galactic center, $E(B-V)\approx3$ at $A(V)>44$\,mag at 550\,nm \citep{2008A&A...488..549P,2011ApJ...737...73F}, an attenuation by a factor of $10^{-18}$. Another prominent feature in measured extinction curves is a ``bump'' in the UV at 217.5\,nm \citep{1965ApJ...142.1683S,1969ApJ...157L.125S}, where extinction is about an order of magnitude higher. It is attributed to organic carbon and amorphous silicates present in the grains \citep{2005Sci...307..244B}. Other features are the water ice absorption at $3.1\,\mu$m and the 10 and $18\,\mu$m silicate absorption.

While higher frequencies have higher channel capacities for coherent waves, and allow for tighter beams (at a given telescope size), they also generally suffer from higher extinction between UV and IR. To analyze this trade-off (section~\ref{sec_ext}), we use the synthetic extinction curve presented in \citet{2003ARA&A..41..241D,2003ApJ...598.1017D,2003ApJ...598.1026D} which covers wavelengths from 1\,cm (30\,GHz) to 1\,{\r{A} (12.4\,keV). We scale this curve for different distances using $A(V)=1.8$\,mag per kpc in the galactic plane \citep{2003dge..conf.....W}, equivalent to $E(B-V)=0.28$\,mag per kpc \citep{2002A&A...389..871D}. For the highest extinction values towards the galactic center, where $E(B-V)\approx3$, we use measurements for the optical and IR \citep{2011ApJ...737...73F} and the UV \citep{2004ApJ...616..912V,2016ApJ...828...69M} and interpolate in between individual data points with a spline. This extinction curves covers the wavelength range from $0.1-27\,\mu$m. While extinction is typically given in astronomical magnitudes, we convert these to the fraction of photons received over distance ($S_{\rm E}$), and show examples in Figure~\ref{extinction}.

\begin{figure*}
\includegraphics[width=.5\linewidth]{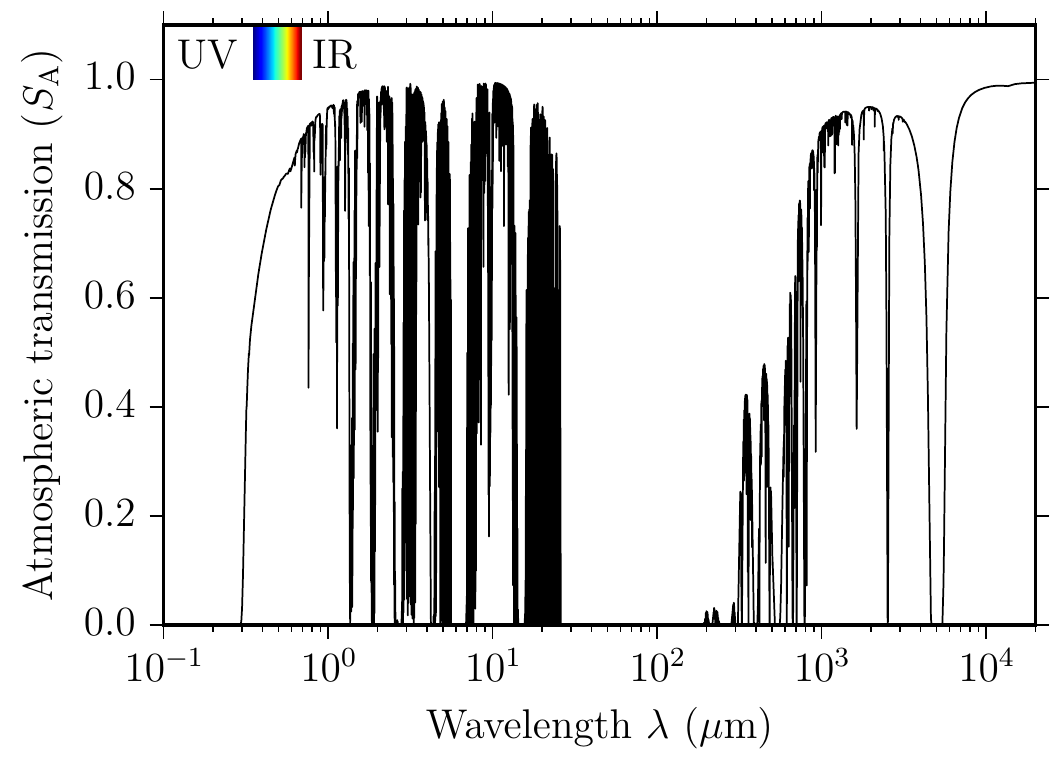}
\includegraphics[width=.5\linewidth]{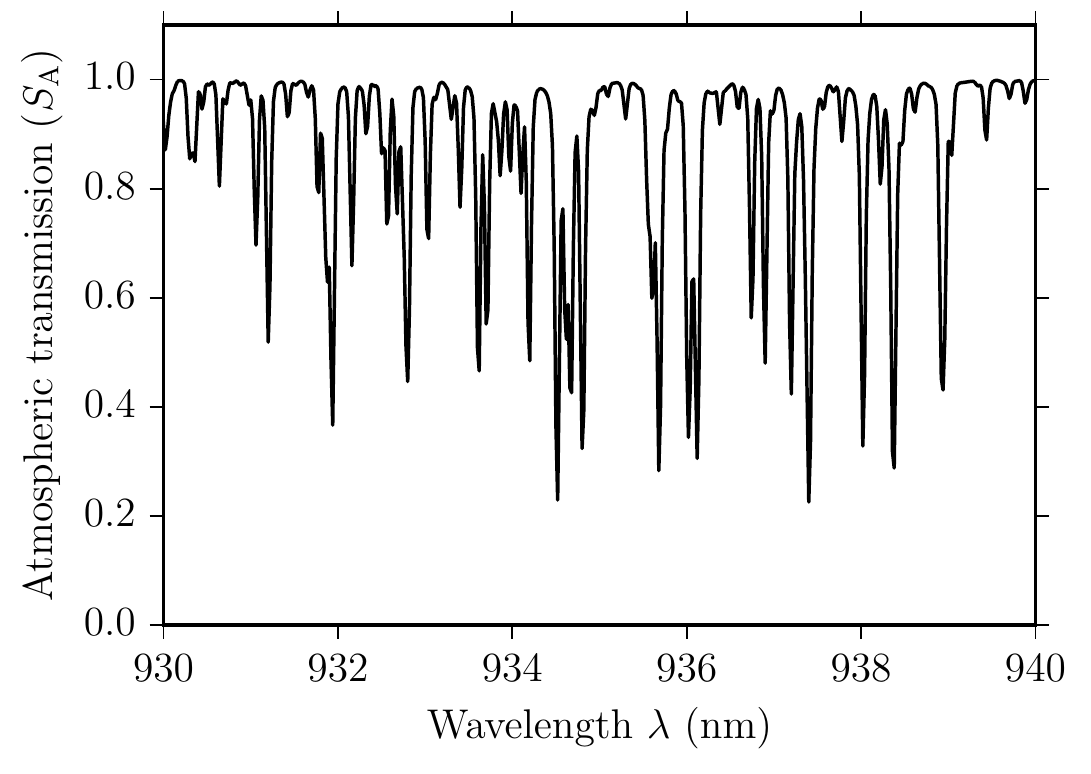}
\caption{\label{figure_atmo}Left: Surviving fraction of photons after atmospheric transmission ($S_{\rm A}$) as a function of wavelength. Short-ward of UV (291\,nm), transmission remains at zero. Data is for Mauna Kea in best (20-percentile) conditions. Right: Zoom into the IR with fluctuations from 0.2 to unity transmission with typical line widths of $2\,{\rm \r{A}}=0.2$\,nm.}
\end{figure*}

\subsection{Loss of photons from atmospheric transmission}
\label{atmo}
The earth is surrounded by an atmosphere \citep{1842RSPT..132..225F}, which is essential for almost all life on this planet \citep{2007Sci...315...92C}, but of the greatest annoyance for almost all astronomers \citep{1950PASP...62..133K}. For a space telescope there is no loss of photons from a surrounding cloud of gas, dust and water, so that the surviving fractions of photons is $S_{\rm A}=1$. On earth, atmospheric transmission depends on the wavelength and varying characteristics, such as the content of water vapor in the air. As an example, we use a transmission curve $S_{\rm A}(\lambda)$ for Mauna Kea with a water vapor column of 1\,mm, which represents excellent observing conditions, and occurs in the 20\% of the best nights of an average year \citep{1992nstc.rept.....L,2009JGRD..11418105G}. This curve covers the wavelength range of 200\,nm--10\,cm (3\,GHz). Figure~\ref{figure_atmo} shows the part up to 10\,mm (30\,GHz), after which transmission reaches near unity.

Transmission is zero for all practical purposes for wavelengths below 291\,nm, above 20\,m, and between $30-200$\,$\mu$m. In the optical and infrared, transmission is highly variable due to numerous absorption lines from water, carbon dioxide, ozone and other gases. When communicating with photons in a narrow (nm) bandwidth, as is common with lasers, the exact wavelength must be chosen carefully, because transmission fluctuates rapidly. For example, $S_{\rm A}=0.98$ at $\lambda=934.36$\,nm, but $S_{\rm A}=0.22$ at $\lambda=934.52$\,nm, a spectral distance of only 0.16\,nm. Under good atmospheric conditions, transmission can be close to unity for many wavelengths in the optical and near- to mid-infrared.

For brevity, we neglect other atmospheric effects such as scattering and turbulence \citep[``seeing'',][]{1995ApOpt..34.5461C} which is a variation of the optical refractive index and enlarges the point spread function of the telescope, if not corrected for with adaptive optics \citep{1998aoat.book.....H}.

\begin{figure}
\includegraphics[width=\linewidth]{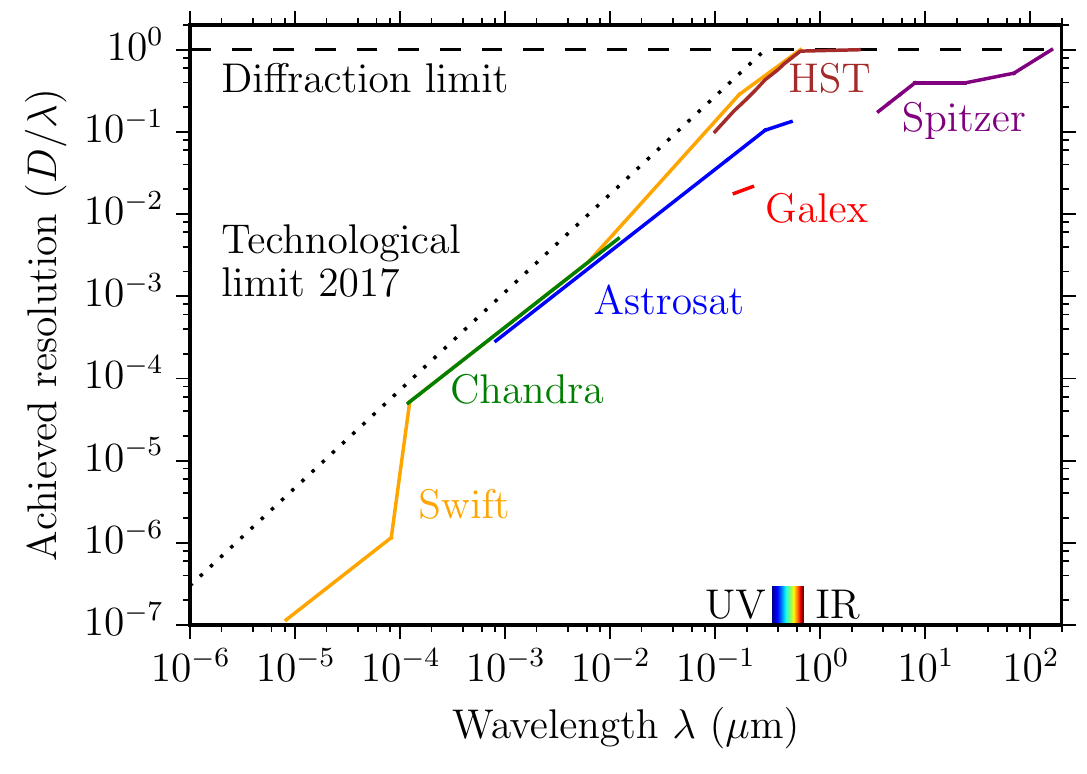}
\caption{\label{figure_lambda_D}Technologically achieved resolution for space telescopes (earth 2017) as a function of wavelength. Focusing high-energy waves is increasingly difficult.}
\end{figure}

\subsection{Technological limits of telescopes}
\label{sec_tech_limit}
The angular beam size is limited to $Q_{\rm R}\geq1.22$ (Rayleigh limit), or $Q_{\rm L}\geq1$ for a laserbeam. Technology may place a stricter limit. We have examined the angular resolution of current (earth 2017) space telescopes for different wavelengths. As can be seen in Figure~\ref{figure_lambda_D}, $Q_{\rm real}/Q_{\rm R}$ is an exponential function for wavelengths $\lambda < 300$\,nm, indicating the technological difficulty to focus wavelengths in UV and shorter.

For diffraction-limited telescope mirrors, the polished surfaces need to have surface smoothness $< \lambda/4$ \citep{1935lett.book.....D}, which makes the production of telescopes for UV, X-ray and $\gamma$-ray increasingly difficult. Additionally, the refractive index of all known materials is close to 1 at high (keV) energies, making it difficult to focus photons efficiently and avoid absorption \citep{2008PhyU...51...57A}.

With today's technology, resolution in the milli-arcsec regime is possible at optical wavelengths, but X-rays are limited to angular resolutions of 20\,arcsec \citep{2014SPIE.9151E..2WS}, a difference of 4 orders of magnitude. For example, the Swift X-Ray satellite has an angular resolution of 18 arcsec at $\lambda=1$\,nm (1.5\,keV) from a 30\,cm aperture \citep{2005SSRv..120..165B}, while the diffraction limit would be $1.22\lambda/D=8\times10^{-4}$\,arcsec, so that $Q_{\rm real}/Q_{\rm R} = 4\times10^{-5}$. Technology is believed to eventually achieve sub-arcsec resolution at X-rays, but at the expense of large designs, with focal lengths of $10^5$\,km \citep{2004SPIE.5168..411G}.

\subsection{Technological limits of the receiver}
\label{tech_limit}
Photon energy depends on wavelength, $E=hc/ \lambda$, which should make it easier to detect higher energy photons in theory. In practice, single photon detection with high quantum efficiency is possible throughout a wide range of wavelengths, from X-Rays \citep{2013MedPh..40d1913T,2015MedPh..42..491T} to microwaves \citep{2012PhRvB..86q4506P,2015arXiv151206939W}. Interestingly, even the human eye can detect single photons in the visible light \citep{2016NatCo...712172T}.

We will neglect a possible wavelength-dependence in quantum efficiency of photon detectors in this paper. This is supported by the much stronger influence from technological limits in focusing beams (section~\ref{sec_tech_limit}), and the influence of interstellar extinction on photon throughput (section~\ref{loss_ext}), so that detector differences (of a few percent) will be negligible for most practical cases.

\begin{figure}
\includegraphics[width=\linewidth]{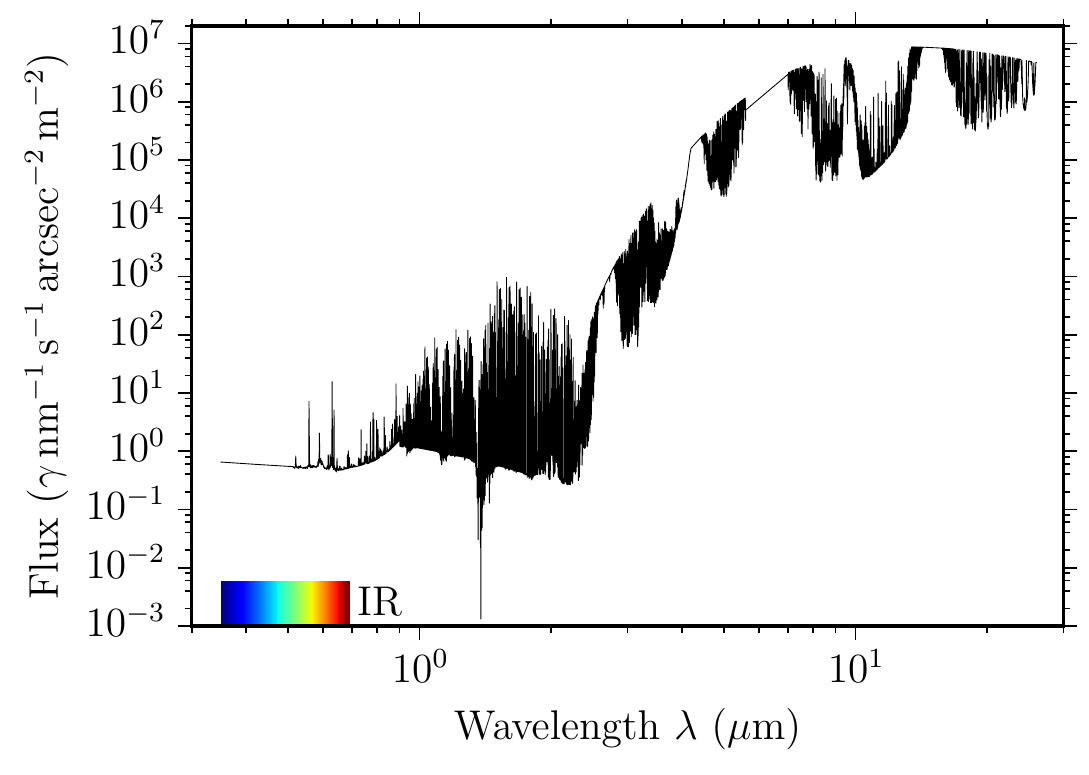}
\caption{\label{figure_atmo_glow}Atmospheric sky background on Mauna Kea as a function of wavelength.}
\end{figure}

\section{Noise}
\label{noise}
Noise sources can be astrophysical (scattering of the signal, background light) or instrumental (shot noise and read noise). For ground-based telescopes, the total noise has been measured (section~\ref{sky}), for space-based telescopes, it will be discussed in sections 4.2--4.6.

\subsection{Atmospheric sky background}
\label{sky}
For a telescope located on earth, the total sky background which enters as noise into the receiver can be measured by observing a (maximally) empty sky area. Naturally, it includes all sources: terrestrial, solar system, and interstellar.

Precise raw sky emission data is available for many observatory sites, and as in section~\ref{atmo} we use Mauna Kea as an example. The measurements are for the sky background only and do not include the emission from a telescope or sensor (which has been subtracted out). The data were manufactured with a synthetic sky transmission \citep{1992nstc.rept.....L} subtracted from unity. This gives an emissivity which is then multiplied by a blackbody function with a temperature of 273\,K \citep{2009JGRD..11418105G}. The authors added emission spectra based on observations from Mauna Kea, and the dark sky continuum mostly from zodiacal light. Finally, the curve has been scaled to produce 18.2\,mag\,arcsec$^{-2}$ in the H band, as observed on Mauna Kea by \citet{1993PASP..105..940M}. The resolution of the final data product is 0.1\,nm\footnote{Data files from \url{http://www.gemini.edu/sciops/telescopes-and-sites/observing-condition-constraints/optical-sky-background}}.

These values are in agreement with measurements from the darkest observatory sites on earth, which have an optical sky background minimum of $22$\,mag\,arcsec$^{-2}$ \citep{2008ASPC..400..152S}, corresponding to an optical flux of a few \,$\gamma$\, arcsec$^{-2}$\,sec$^{-1}$ from unresolved sky sources, air glow, and zodiacal light.

The sky background at Mauna Kea is shown in Figure~\ref{figure_atmo_glow} and covers the band from 300\,nm--30\,$\mu$m. Similarly to the transmission (section~\ref{atmo}), background levels vary by up to 3 orders of magnitude over few nm. Generally, the flux is $\approx \gamma$\,nm$^{-1}$\,arcsec$^{-2}$\,m$^{-2}$ in the optical and NIR, with a steep increases for $\lambda>2.5$\,$\mu$m, and reaches $10^7 \gamma$\,nm$^{-1}$\,arcsec$^{-2}$\,m$^{-2}$ at 10\,$\mu$m.

This indicates that earth-based interstellar communication is favorable for $\lambda<2.5$\,$\mu$m. For telescopes on other planets, we would need to know precisely the exoplanet atmospheres, exozodiacal dust, etc. which may result in a different noise structure; a detailed discussion is beyond the scope of this paper.

\begin{figure}
\includegraphics[width=\linewidth]{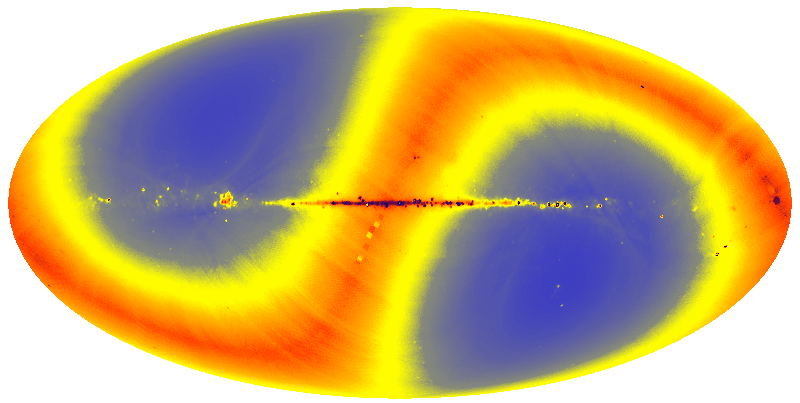}
\caption{\label{zodi2}All-sky map at 12\,$\mu$m taken by the COBE satellite \citep{1992ApJ...397..420B,1998ApJ...508...44K}. The horizontal line is the galactic plane, the S-shaped band represents the solar system ecliptic, where zodiacal light is $>100\times$ higher than near the ecliptic poles \citep[blue colors,][]{1980A&A....84..277L}.}
\end{figure}

\subsection{Background light from zodiacal light}
\label{noise_zodi}
Space telescopes are not affected by the strong atmospheric light. However, they still collect undesired photons. The strongest source is sunlight which is scattered off of dust grains in the solar system, an effect called zodiacal light. In the ecliptic plane, it can be as bright as $1.5\times10^{-6}$\,ergs\,s$^{-1}$cm$^{-2}$\r{A}$^{-1}$. It is faintest at heliocentric longitude 130$^{\circ}-170^{\circ}$ away from the sun because of larger scattering angles, and at low ecliptic latitudes $<30^{\circ}$ because of the minimum in the interplanetary dust column density at levels $<10^{-7}$\,ergs\,s$^{-1}$cm$^{-2}$\r{A}$^{-1}$ \citep{2002ApJ...571...56B}. The scattering strength only weakly depends on wavelength and closely resembles the solar spectrum between 150\,nm and 10\,$\mu$m \citep{1981A&A...103..177L,1995Icar..115..199M}.

These levels contribute a flux of order 3\,$\gamma$\,nm$^{-1}$\, arcsec$^{-2}$\,m$^{-2}$ at 1\,$\mu$m in the ecliptic, and decrease to 0.1 (0.03) photons at latitude 45$^{\circ}$ (90$^{\circ}$). We show an all-sky map in Figure~\ref{zodi2} which makes it clear that the source's location on the sky is important, in addition to the wavelength.

\begin{figure}
\includegraphics[width=\linewidth]{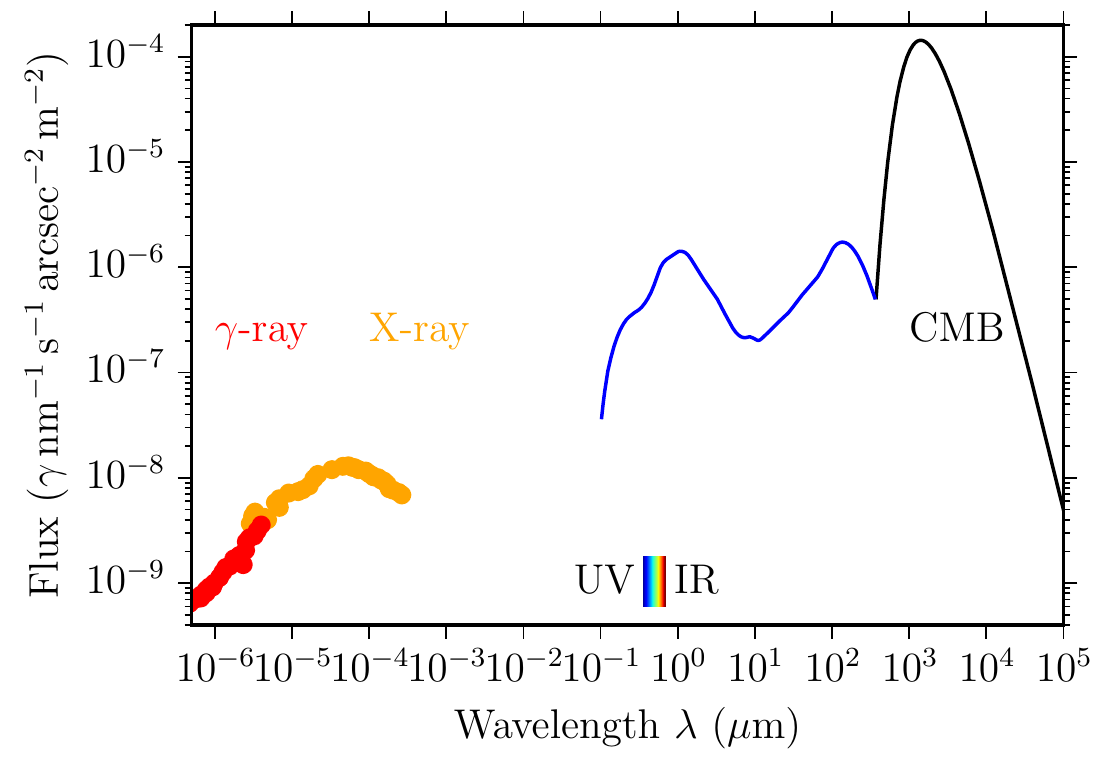}
\caption{\label{figure_background_sources}Intensity of the extragalactic background after removal of the zodiacal light foreground (which is strongest in the visible and IR). The peak in the optical is from nuclear fusion, the peak in the FIR from re-radiated dust. The UV/soft X-ray background at a wavelength of 10--100\,nm is unknown. Data from \citet{1998ASPC..139...17L,2016RSOS....350555C,2016ApJ...827....6S}.}
\end{figure}

\subsection{Background light from galactic and extragalactic sources}
\label{noise_atmo}
The Galactic light comes from stars, starlight scattered from interstellar dust, and line emission from the warm interstellar medium. Its levels are of order $10^{-9}$\,ergs\,s$^{-1}$cm$^{-2}$\r{A}$^{-1}$ between 200\,nm and 1\,$\mu$m.

The mean flux of the optical extragalactic background light has been measured as $4.0\pm2.5$, $2.7\pm1.4$ and $2.2\pm1.0 \times 10^{-9}$\,ergs\,s$^{-1}$cm$^{-2}$\r{A}$^{-1}$ at wavelengths of 300\,nm, 550\,nm and 800\,nm \citep{2002ApJ...571...56B}.

Compared to the zodiacal light, these sources are weaker by two orders of magnitude and are only relevant if the source is near the ecliptic poles, where zodi is smallest; and for wavelengths outside the zodi-band of $\approx0.3-300$\,$\mu$m.

\subsection{Scintillation and scattering of photons}
\label{scatter}
Extinction causes not only a loss of photons from absorption, but also scattering. The latter reduces the ``prompt'' pulse height and produces an exponential tail \citep{2000ASPC..213..545H}.

Scatter broadening is well known from pulsars and magnetars. As an extreme example, magnetars close (0.1\,pc) to the galactic center with dispersion measures $\text{DM}=1778$\,pc\,cm$^{-3}$ have their pulses broadened to $1.423\pm0.32$\,s at 1.2\,GHz, and $0.2\pm0.07$\,ms at 18.95\,GHz, following a power law with index $-2.8$ \citep{2014ApJ...780L...3S}. A single pulse which is broadened to a width of one second results in a very low data rate of order bit/s. Extrapolating with the power law indicates that nanosecond pulse widths ($10^{-9}$\,s) can be expected for frequencies $>500$\,GHz ($\lambda<0.6$\,mm), and the broadening should become shorter than the wavelength at $\lambda \approx \mu$m.

For these higher frequencies, the amplitude level of the scattering tail, and its length, is unknown in practice. Limits from the Crab pulsar show no detectable scattering tail at UV and optical wavelengths for an optical millisecond pulse width and $E(B-V)=0.52$ \citep{2000ApJ...537..861S}, consistent with the power law scaling from radio observations. These results indicate that the impact of extinction is mainly on the absorption for frequencies $>500$\,GHz ($<0.6$\,mm), and not on pulse broadening. Therefore, we neglect this effect in our calculations, but suggest further research in this area.

\subsection{Background light from the target star and celestial bodies}
\label{star}
On the direct path, even modest-sized telescopes receive a relevant number of photons from nearby stars. For example, $5\times10^{10}\,\gamma$\,sec$^{-1}$\,m$^{-2}$ from $\alpha$\,Cen~A \citep[distance 1.3 pc,][$L= 1.522 L_{\odot}$]{2016A&A...594A.107K}. From Proxima Centauri ($L=1.38\times10^{-4} L_{\odot}$), it is $4.25\times10^{6}\,\gamma$\,sec$^{-1}$\,m$^{-2}$, or $3.5\times10^9$ ($3.5\times10^{5}$)\,$\gamma$\,sec$^{-1}$\,m$^{-2}$ from a sun-like star in a distance of 10 LY (1000 LY). A coronograph or occulter could be used to block a significant part of this flux \citep[$10^{-9}$,][]{2006ApJS..167...81G,2015RAA....15..453L}. Additionally, a filter with a small band-pass, e.g. 1\,nm, would reduce the flux further by $>10^3$. A good angular resolution of the receiving telescope would be helpful to separate the transmitter from the nearby target star. For comparison, a probe at a distance of 1\,au from the star $\alpha$\,Cen~A would appear at an angular separation of 1.42\,arcsec as seen from earth, resolvable even with small telescopes, assuming sufficient contrast.

The flux levels from reflected exoplanet light and exozodiacal dust is many orders of magnitude fainter than from the flux in the home solar system, and can thus be neglected.

\subsection{Instrumental noise}
\label{instrumental_noise}
Apart from a loss of photons from imperfect reflection or transmission in the receiver, the conversion from photons to electrons (e.g. with CCDs or photomultiplier tubes, which are analogue devices) causes a small, but nonzero amount of noise.

Even a perfect instrument will produce some noise. Fundamentally, this originates from the fact that photons and electrons are quantized \citep{1905AnP...322..132E}, so that only a finite number can be counted in a given time. This phenomenon is the shot noise \citep{1918AnP...362..541S}, and is correlated with the brightness of the target.

\begin{figure*}
\includegraphics[width=.5\linewidth]{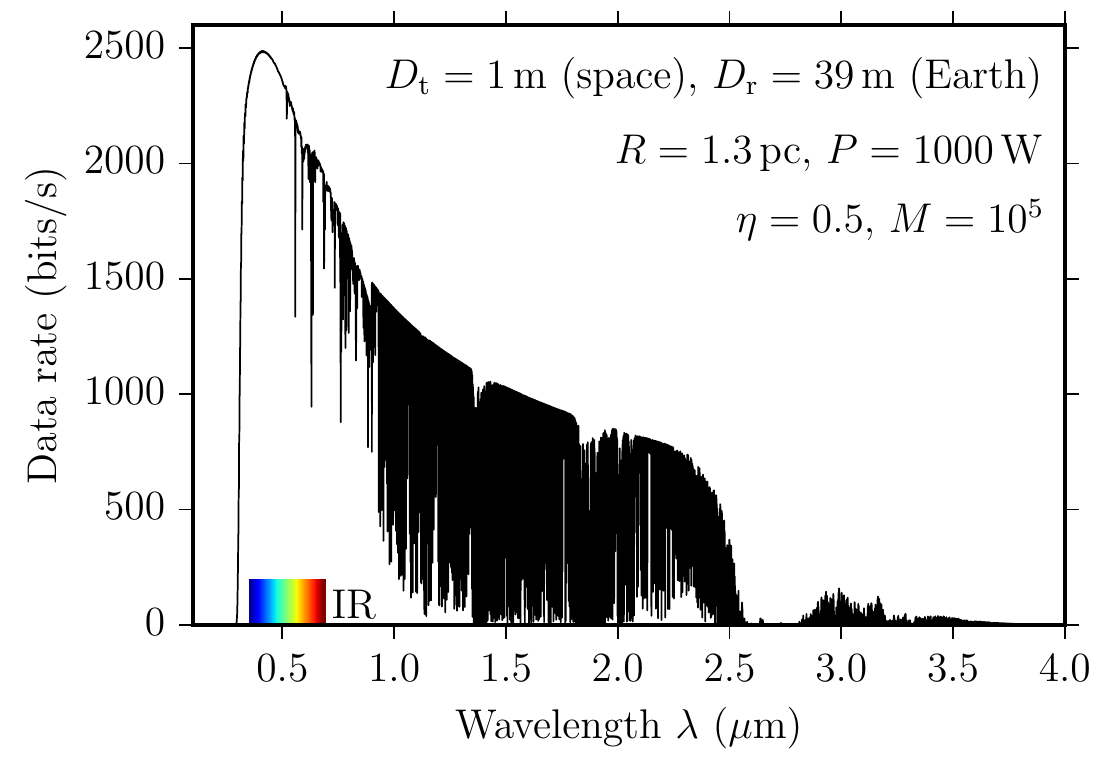}
\includegraphics[width=.5\linewidth]{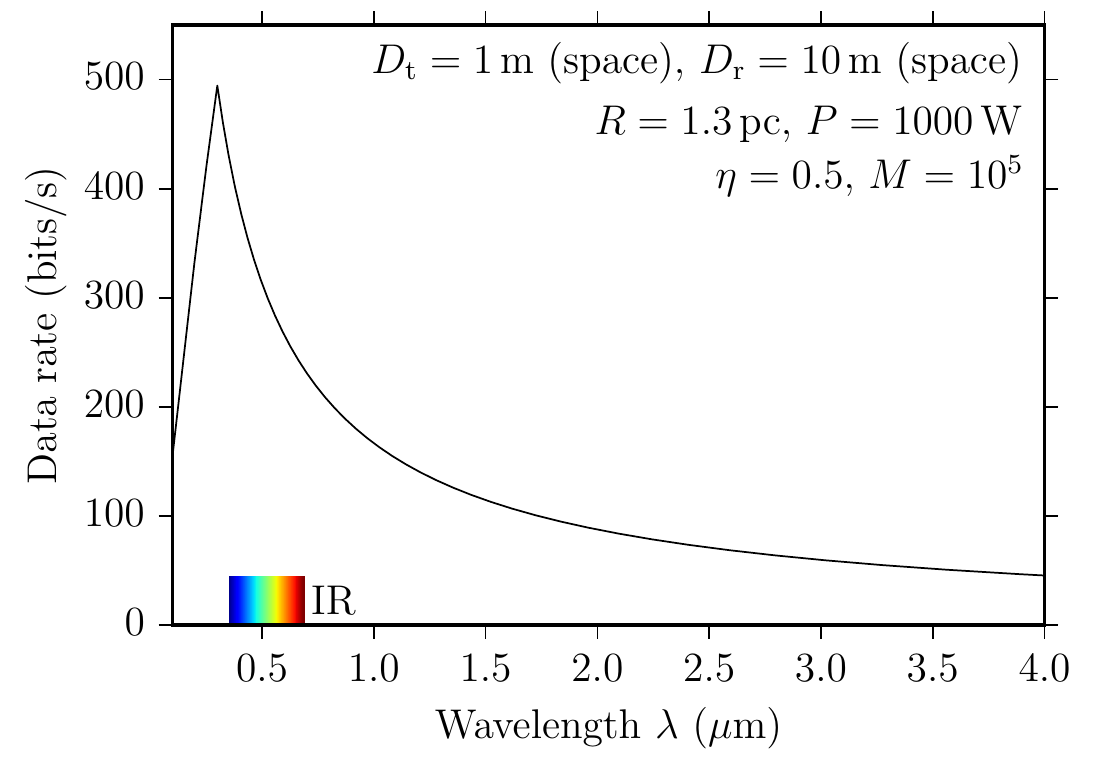}
\caption{\label{proxima}Data rate to a probe at Proxima, as a function of wavelength, for the listed parameters. Left: Receiver on earth peaks for $\lambda=429.6$\,nm. Right: Receiver in space peaks at 300\,nm. See text for discussion.}
\end{figure*}

\section{Results}
\label{results}

\subsection{A Starshot-like probe at $\alpha$\,Centauri}
We will now calculate exemplary quantitative data rates. Our default example is to maximize data rates with a probe at $\alpha$\,Cen ($d=1.3$\,pc) and examine the influence of the variables presented in the previous section. Our standard example probe uses a telescope with a circular aperture $D_{\rm t}=1$\,m, through which it transmits with a power of $P=1,000$\,W. The telescope quality $Q_{\rm R}\approx1.22$ for $\lambda>300$\,nm is of current (earth 2017) technology, and positioned in space. The hypothetical receiver has $D_{\rm r}=39$\,m, comparable to the upcoming generation of ``extremely large telescopes'' (E-ELTs). It must be located in the southern hemisphere, e.g. in Chile, because $\alpha$\,Cen is not observable from Mauna Kea's northern latitude which served as an example in previous sections. The total receiver efficiency is $\eta=0.5$. It uses $N=10^5$ modes, which could be done with a $R=100,000$ spectrograph, $10^5$ time slots, or a combination of both. The atmospheric sky background represents very good (20-percentile) conditions as described in section~\ref{sky}.

From the transmitted $P=1,000$\,W ($2.2\times10^{21}$\,$\gamma$\,s$^{-1}$ at $\lambda=429.6$\,nm), the theoretical flux near earth after free-space loss is $1,860$\,$\gamma$\,s$^{-1}$ in the receiver aperture. Interstellar extinction for this wavelength and distance is $\approx0.3$\%, causing a loss of 6 photons, or a reduction to $1,854$\,$\gamma$\,s$^{-1}$. Sky transparency is 0.74, so that $1,369$\,$\gamma$\,s$^{-1}$ survive. This is the signal before receiver efficiency.

Regarding the total sky background, we assume that the filter width at the receiver has a bandpass of 1\,nm, and the on-sky resolution is 1\,arcsec. We neglect the photon flux from $\alpha$\,Cen as it can be effectively suppressed (section~\ref{star}), and is then negligible compared to the atmospheric background of 0.6\,$\gamma$\,nm$^{-1}$\,s\,arcsec$^{-2}$\,m$^{-2}$ (section~\ref{atmo}), resulting in 702 noise photons per second in the telescope. We will discuss the case of blended sources (probe and star) in section~\ref{sec:blend}. We also neglect the noise flux from the receiver itself. Following Eq.~\ref{thermal_holevo}, the Holevo bound with our noise is then 1.81 bits per photon. This includes the receiver efficiency of $\eta=0.5$.

We can now multiply the received photons with the encoding limit and estimate $1,369$\,$\gamma$\,s$^{-1}$ at $1.81$\,bits\,$\gamma^{-1}$ $=2480$\,bits/s. This is also the peak value at $\lambda=429.6$\,nm in Figure~\ref{proxima} (left), indicating that any other wavelength decreases the effective data rate. In practice, this is an upper bound; realistic data rates including sensor noise, margin for error, etc. will yield smaller data rates by a factor of a few.

The cut-off for $\lambda<290$\,nm comes from the atmospheric intransparency (Figure~\ref{figure_atmo}). The decline in data rate towards longer wavelengths comes from two effects: the decrease in telescope focusing (section~\ref{sec_tech_limit}), and increasing atmospheric noise (Figure~\ref{figure_atmo_glow}). Individual atmospheric absorption lines can be clearly seen which should be avoided for communication (Figures~\ref{figure_atmo},~\ref{figure_atmo_glow}).

\begin{figure}
\includegraphics[width=\linewidth]{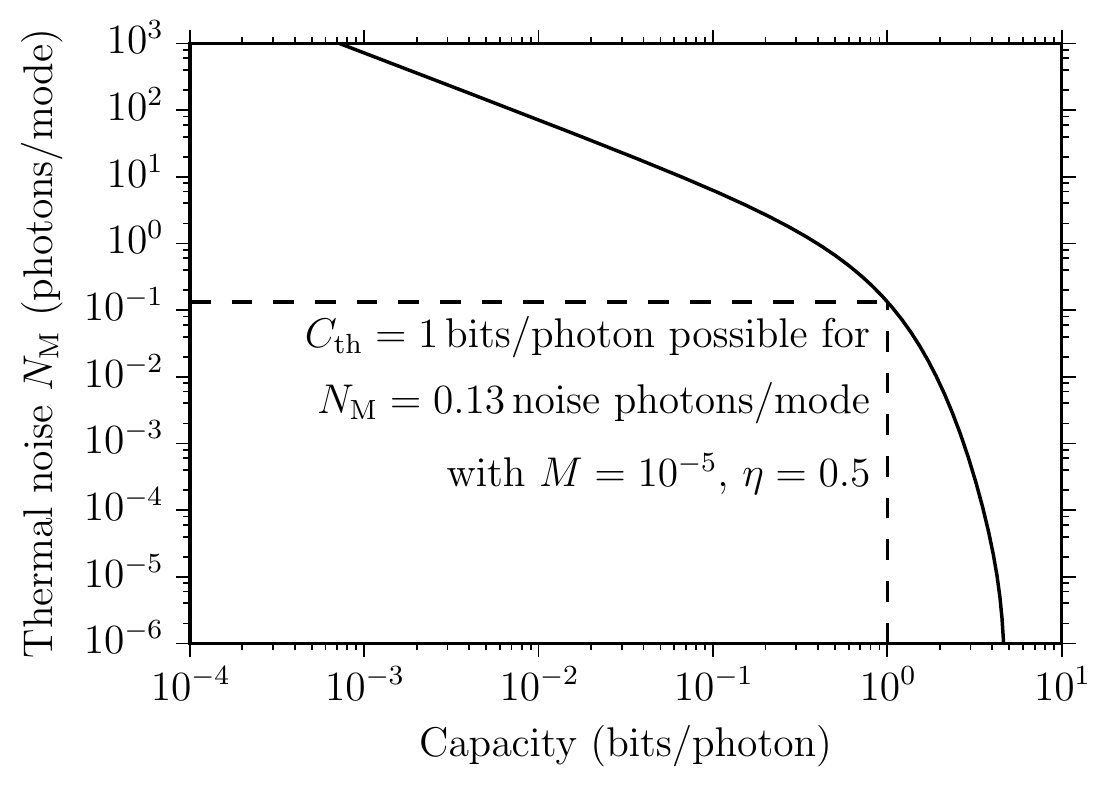}
\caption{\label{hnoise}Capacity $C_{\rm th}$ (in bits/photon) is a logarithmic function of thermal noise $N_{\rm M}$ (in photons per mode).}
\end{figure}

\subsection{Space-based receiver}
For the space-based analysis, we restrict the receiver size to $D_{\rm r}=10$\,m to make it more realistic for the current technological level. The optimal wavelength is now $\lambda \approx 300$\,nm which is limited by the telescope quality (Figure~\ref{figure_lambda_D}). Noise levels are dominated by zodiacal light; $\alpha$\,Cen is $42^{\circ}$ from the ecliptic, resulting in noise levels of $\approx0.1$\,$\gamma$\,nm$^{-1}$\,s\,arcsec$^{-2}$\,m$^{-2}$ and a higher capacity of 2.83 bits per photon. The signal decreases to 174\,$\gamma$\,$^{-1}$ for a maximum data rate of 494\,bits/s.

\begin{figure*}
\includegraphics[width=.5279\linewidth]{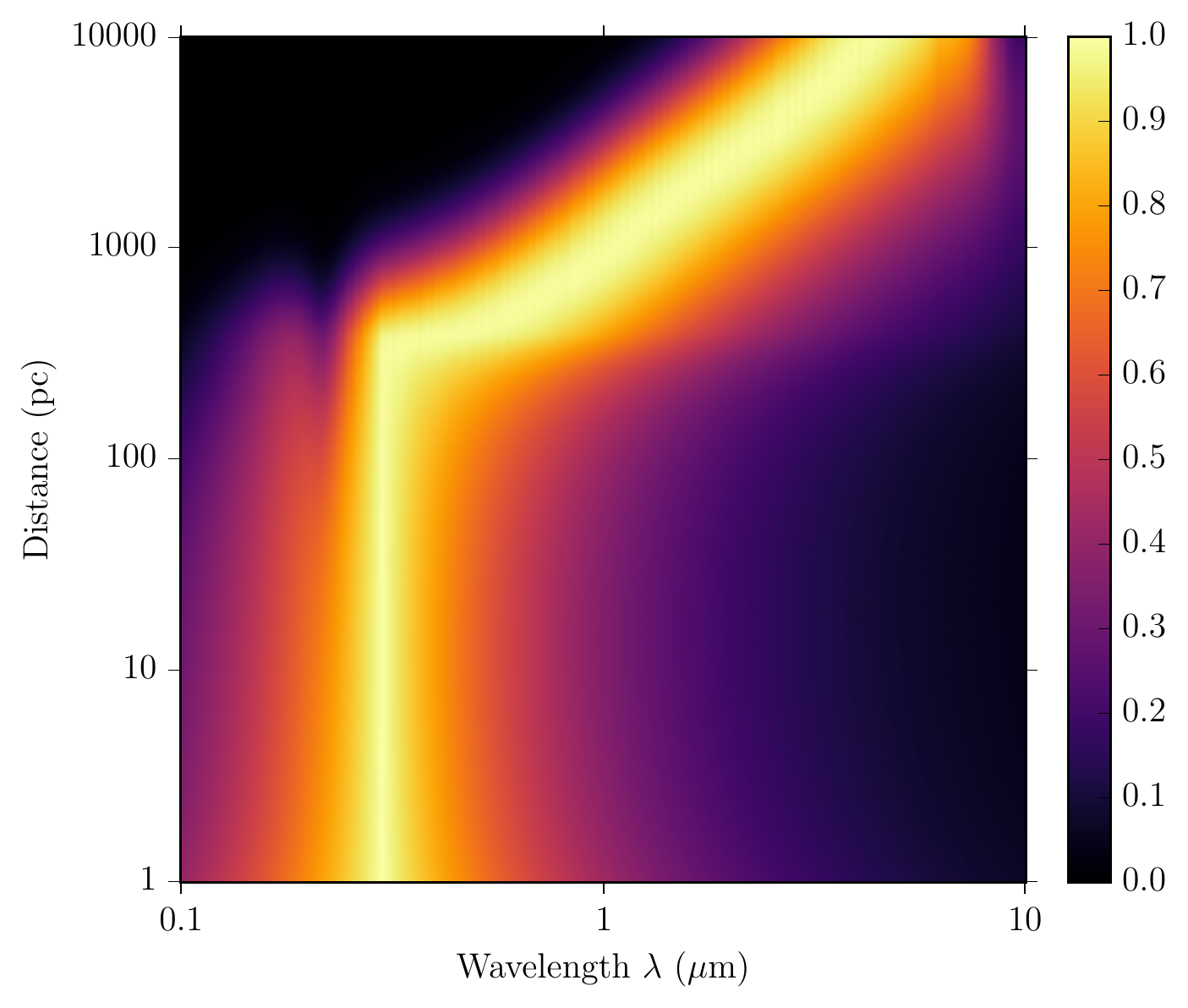}
\includegraphics[width=.4721\linewidth]{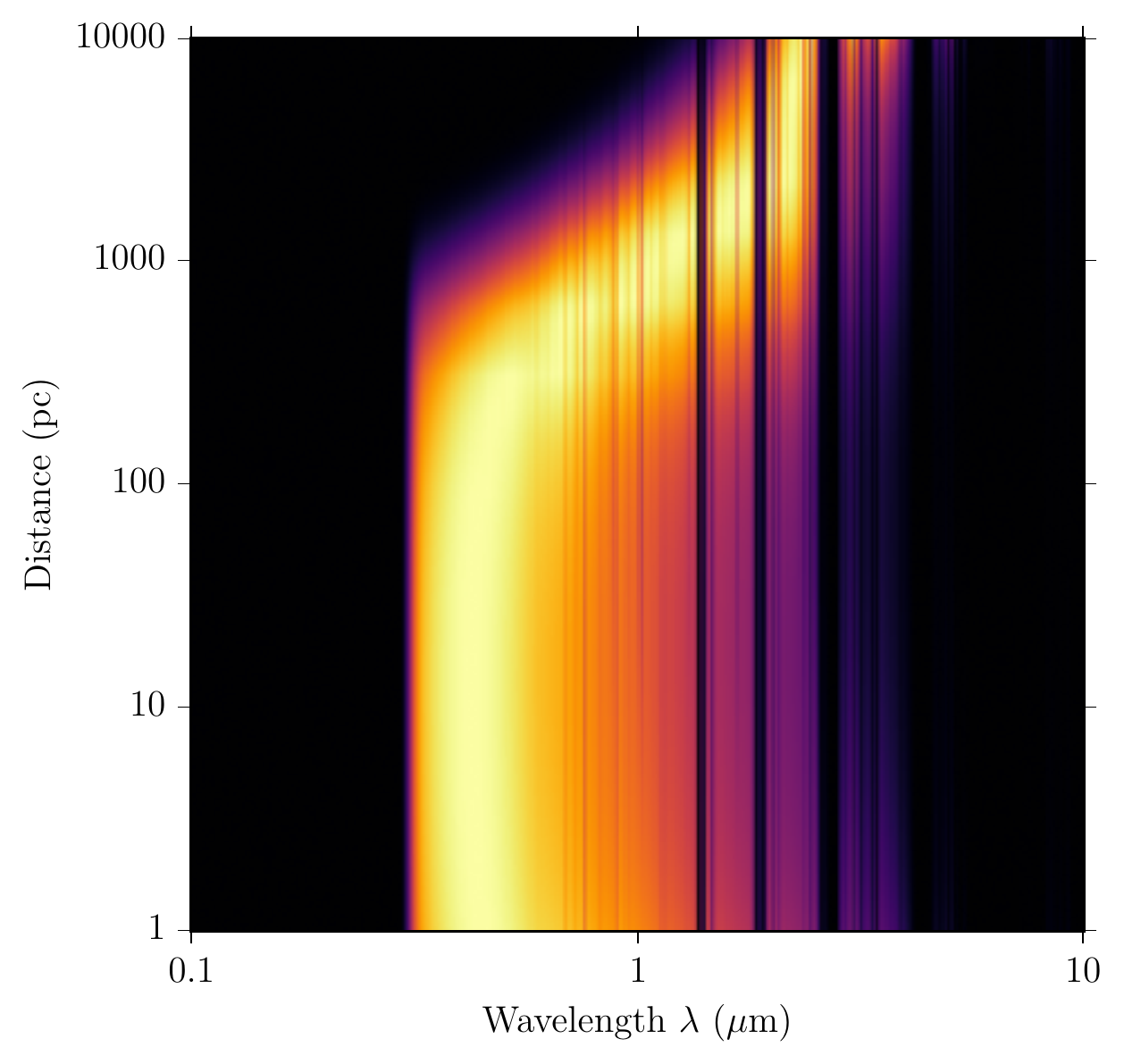}
\caption{\label{fig_photons}Best frequency (brightest color) as a function of distance. The influence of free space loss has been subtracted out as it would have overpowered all other parameters. Left: Space-based telescope. Right: earth-based, including atmospheric transmission. The optimal wavelength is close to $0.3$\,$\mu$m for distances $<200$\,pc and increases to the mid-IR for larger distances.}
\end{figure*}

\subsection{Power}
At first approximation, data rate is a linear function of power, $\text{DSR}\textsubscript{$\gamma$} \propto P$. This holds for constant capacity $C_{\rm th}$ which however depends on the ratio of signal to noise, and thus decreases for decreasing signal. The effect is small for $S \gg N$ but becomes very considerable for $N>S$. As shown in Figure~\ref{hnoise}, a capacity $C_{\rm th}=1$\,bits per photon is possible for $N_{\rm M} \leqq 0.13$ (noise photons per mode) in our standard example using $M=10^{-5}$ modes and receiver efficiency $\eta=0.5$. Capacity is a logarithmic function of SNR, and the sweet spot appears between 0.1--5 bits/photon, which is achievable for $10^{-6}<N_{\rm M}<10$ assuming $10^{5}$ modes and $\eta=0.5$.

\subsection{Transmitter size}
The transmitter size for a circular aperture scales as $\text{DSR}\textsubscript{$\gamma$} \propto D_{\rm t}^2$ assuming no technological limitations, which we identify as possible for current (earth 2017) technology at $\lambda>300$\,nm. Increasing the dish size to focus optical lasers is thus very beneficial for the data rate, and it is recommended to make the aperture as large and high-quality as possible.

\subsection{Receiver size}
The receiver size for a circular aperture scales as $\text{DSR}\textsubscript{$\gamma$} \propto D_{\rm r}^2$, and we here relax the technological limitations: imperfect focusing will still collect all photons (signal), but collect more noise due to the larger beam width; the total effect is however much smaller. For a real application, this additional noise factor can be modeled.

\subsection{Interstellar Extinction}
\label{sec_ext}
Extinction is largely irrelevant for the shortest interstellar distances, $<1$\% in the optical to $\alpha$\,Cen. Outside of the Lyman continuum ($\approx50<\lambda<91.2$\,nm), any frequency is equally suitable. The situation changes significantly for distances $>200$\,pc, where optical extinction is $>0.5$ (compare Figure~\ref{extinction}). To examine the optimal choice of wavelength versus distance due to extinction, we have plotted the normalized photon rate in Figure~\ref{fig_photons}, and subtracted out the free-space loss. The optimal wavelength for space-based communication is limited by technology at 300\,nm out to 200\,pc, and increases to 3\,$\mu$m for the longest paths in the galaxy. For an earth-based receiver, the lower limit is 420\,nm due to limited atmospheric transmission, and special care must be taken not to select a narrow absorption line.

In this calculation, we assumed uniform extinction of $A(V)=1.8$\,mag per kpc in the galactic plane \citep{2003dge..conf.....W}. In reality, however, the situation is much more complex. Extinction in the galactic plane can vary on small scales (because of individual molecular clouds), and on large (degree) scales \citep{2016ApJ...821...78S}. Galactic communication with maximized data rates will require a precise measurement along each line of sight (communication path) to choose the best wavelength. If a civilization, or a club of civilizations, prefers to choose a single frequency for all distances, it will be at $\approx3\,\mu$m. Then, long distance communication is near optimal (it would be prohibitive at shorter wavelengths), while data rates for short-distance communication are smaller by a factor of a few compared to individual optima.

\begin{figure*}
\includegraphics[width=\linewidth]{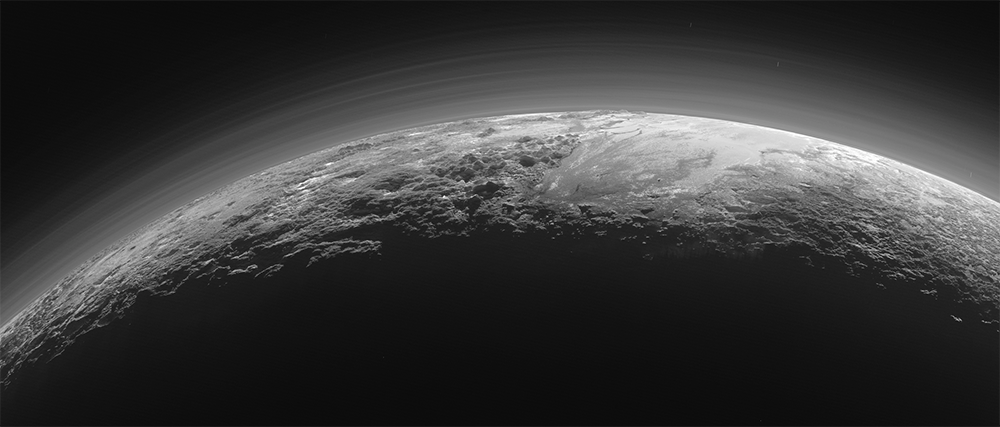}
\caption{\label{pluto}Pluto image taken by ``New Horizons'' with a compressed (lossy) data volume of $\approx400$\,kbits for the shown quality.}
\end{figure*}

\section{Discussion}

\subsection{Assessment of achievable data rates}
Achievable data rates are of order kbits/s per KW for a meter-sized probe at $\alpha$\,Cen. For comparison, the NASA probe ``New Horizons'' achieved a data rate of 1\,kbits/s at $P=13$\,W from Pluto, and transmitted a total of 50\,Gbits ($5\times10^{10}$\,bits, buffered) over the course of 15 months. The transfer of an image as shown in Figure~\ref{pluto} with a compressed volume of $\approx400$\,kbits takes 7\,min to transfer at 1\,kbits/s for a $P=1$\,kW at $\alpha$\,Cen, or days (to weeks with problematic SNR) at $P=1$\,W, which might be regarded as acceptable.

Photovoltaic energy is available at a level of kW\,m$^{-2}$ at au distance from the star, so that a probe in orbit (perhaps decelerated by stellar photons, \citet{2017ApJ...835L..32H,2017arXiv170403871H}) has no power issues for transmissions. A fly-by probe at 20\%\,c, however, transverses the au distance in 17 minutes, translating into a data volume of order Mbits\,m$^{-2}$ if the whole time were used for transmission (which is unrealistic, given that the target exoplanet is to be observed). Available photovoltaic energy decreases with the inverse square to the distance from the star, and by integrating over an exemplary trajectory with a closest encounter of 1\,au to the star we can estimate the total collected photovoltaic energy, during the fly-by, of order kWh\,m$^{-2}$. With this energy, perhaps stored on-board and used for later transmission, the probe can send a few Mbits\,m$^{-2}$, i.e. a few high-quality images (Figure~\ref{pluto}). Alternative options would require an onboard energy source.

\subsection{Onboard storage requirements}
The data volume during the fly-by governs the size of the transmission buffer. ``New Horizons'' carried a total of 16 GBs, which contained all the data it recorded during the fly-by, and which was transfered afterwards. The same scheme could be used for a fly-by at $\alpha$\,Cen. If the probe starts to transmit after the observations, and transmits 1\,bit/s (1\,kbit/s) for a total of 10 years remaining lifetime, it can transfer (and needs to store) a total of 31\,Mbits (31\,Gbits). Both are low numbers and can be stored with current (Earth 2017) technology at millimeter sizes and milligram masses.

\subsection{Earth's rotation}
A space-based receiver, for example at a Lagrangian point, can be used near-continuously. Earth-based telescopes, however, suffer from Earth's rotation (daylight) and weather. When ``New Horizons'' encountered Pluto, the entire NASA Deep Space Network was online to ensure there were no breaks in reception. If the communication scheme with $\alpha$\,Cen is the same, a large number of telescopes will be required. We can, however, replace (expensive) 39\,m E-ELTs with a number of smaller telescopes. To replace one E-ELT in terms of aperture, $\approx1,500$\,telescopes with $d=1$\,m are required, or 24 telescopes with $d=8$\,m.

\subsection{Laser line width, orbital motion, beam sizes}
\label{orbital_motion}
Transmitter and receiver are in relative motion, which results in a change of path length, as already noted by \citet{2013arXiv1305.4684M,2015AcAau.107...20M}. If the sender (receiver) is located on a planet which orbits a star, the Doppler shift will cause a shift in the sender (receiver) frequency. For example, earth's equatorial speed is 465.1\,m/s, or a frequency shift of $1.55\times10^{-6}$. This is an order of magnitude smaller than current spectrographs ($R=100,000$), but larger than typical high-power laser line-widths (350\,MHz, or $6\times10^{-6}$, \citet{1999ApOpt..38.6347D}) by a factor of a few. Laser line width in the mHz range, although at low ($10^{-12}$\,W) power, have been demonstrated \citep{2009PhRvL.102p3601M,2012NaPho...6..687K}. For such small line widths, the shift would need to be modeled and compensated. Regarding noise per mode (atmospheric, zodiacal etc.), very narrow line widths are preferred.

Narrow line widths may give rise to additional noise sources, namely instrumental frequency shifts in the sender and/or receiver, or a change in the interstellar scattering geometry, which may also result in non-gaussian noise per sub-channel.

For the closest stars within a few pc, large optical telescopes (10-100\,m) have diffraction-limited (adaptive optics) beam sizes smaller than typical orbits (au) of exoplanets. When using such tight beams in the transmitter, the position of the receiving telescope (e.g. on a planet in motion) needs to be known with high accuracy at the time of arrival of the photons \citep{1992lbsa.conf..637S,2016QuEle..46..966M}.

\subsection{Blend of probe and star}
\label{sec:blend}
In the previous sections, we have neglected the noise flux from the target star. This is justified for sky-projected separations which allow for the use of coronographs, and suppress $10^{-9}$ \citep{2006ApJS..167...81G,2015RAA....15..453L} of the starlight ($4.25\times10^{6}\,\gamma$\,sec$^{-1}$\,m$^{-2}$) at a separation of 1\,au at Proxima Centauri. During most of the flight, the problem is much less severe because of the large proper motion of $\alpha$\,Cen \citep[3.7\,arcsec\,yr$^{-1}$,][]{2016A&A...594A.107K}.

We can estimate data rates for this increased noise level within the Holevo bound for this situation, and get a capacity of order $10^{-5}$\,bits per second per Watt. Such a low data rate is insufficient for the transfer of images or other observational data, but may be sufficient for simple telemetry and onboard health status.

\subsection{Current technological level and photon dimensions}
The Holevo bound assumes the use of a number of modes to encode information into photons. The available modes in photons are their time of arrival (sometimes called phase modulation), their frequency (or color), and their polarization. Realizing $10^{-5}$ photons per mode will require many ($>10^5$) modes to encode the information. This can be done with a combination of color, timing and polarization. Commonly used are time-frequency modulations. The usage of polarized light is less common, but might be beneficial for our case. Starlight is polarized only by a few percent \citep{2002AIPC..609...44F}, so that the use of polarization, which is possible for lasers, can reduce noise levels by a factor of two.

We now examine currently available technology. For the sender, the shortest possible laser pulse length has decreased by 11 orders of magnitude during the last 50 years, from $100\,\mu$s in the free-running laser of \citet{1960Natur.187..493M} to 67 attoseconds \citep[$10^{-18}$\,s,][]{2012OptL...37.3891Z}. For a detailed history of the exponential improvements, see \citet{2004RPPh...67..813A}. While the pulse length is very short, the repetition rate is slower by many orders of magnitude.

The highest data rates are currently found in fiber-optic communication by sending pulses of light through an optical fiber, with a current record of order Tbits/s ($10^{12}$\,bits/s) on one glass fiber \citep{2016NatSR...621278M}. Commercial products are available with data rates $1-3$ orders of magnitude below this value. The industry standard employs 100 channels with a channel spacing of 100 GHz (0.8\,nm) between $1530-1612$\,nm with a typical bandwidth (frequency range) of $186-196$\,THz \citep{ITU2012}. Limiting factors are small bandwidth (82\,nm, or $b=5$\%), the wavelength stability of lasers with thermal changes, signal degradation from nonlinear effects in optical fibers, inter-channel crosstalk and (clock) timing jitter.

On the receiver side, current photon-counting detectors can be relatively fast (timings below $10^{-10}$\,s) and efficient ($>90$\%) with a low dark count rate ($<1$\,c.p.s.), but suffer from longer ($10^{-7}$\,s) reset times \citep{2013NaPho...7..210M}. Classical photomultiplier tubes offer timings (and reset times) of $10^{-9}$\,s \citep{2006NIMPA.563..368D}. Current photon detectors are fast enough to sample 10\,GHz frequencies at the Nyquist limit \citep[$B<f/2$,][]{1928TAIEE..47..617N}. These limits, however, are technological, and further improvements can be expected. The ideal instrument for high-mode communication would be a high-throughput, high-resolution spectrograph with low-noise, high-speed photon counters on each sub-channel.

\subsection{Bi-directional communication}
The focus in this paper was the communication from a distant, small, power-limited probe towards home-base. The opposite way, perhaps to send new instructions, is comparably easier: Home-base has less stringent limits on aperture size and power. Telescope diameters might be larger by 1--2 orders of magnitude, and power by several orders of magnitude. A major issue might be that the probe needs to ``listen'' at the moment the photons arrive, and not spend the time sending, making observations, or in hibernation. A simple solution would be pre-arranged timeslots.

\subsection{Comparison to the literature}
\label{lit}
In his ``Roadmap to Interstellar Flight'', \citet{2016arXiv160401356L} recently approximated the communication flux as (his section 5.6, our notation) $F=D^2 P / (4 d^2\lambda^2)$ which yields an overestimate by $\approx11.7$\% compared to our Eq.~\ref{eq2}.

In their ``Search for nanosecond optical pulses'', \citet{2000ASPC..213..545H} and \citet{2004ApJ...613.1270H} describe the received photon flux as
\begin{equation}
N_{\rm d}= \frac{\pi^2 D^2 D^2 E_{\rm p}}{16 \lambda d^2 h c}
\end{equation}
(their Eq. 2, neglecting extinction; they set $D=D_{\rm r}=D_{\rm t}$). Numerically, this produces a received photon flux which is too high by $\approx3.67\times$.

In their ``Search for Optical Laser Emission'', \citet{2015PASP..127..540T} define the received photon flux in the same form as in our Eq.~\ref{eq2} (their Eq.~5), but with an incorrect divisor of 2, resulting in $4\times$ too many photons received.

The work by \citet{1996SPIE.2704...61H} discusses beam widths and frequencies of interstellar laser communications, but neglects extinction, and consequently proposes laser communication in the Lyman H$\alpha$ line at 126\,nm over distances of $3,000$\,LY, which is impossible because of very high UV extinction (Figure~\ref{extinction}).

A more traditional interstellar radio communication design from $\alpha$\,Cen has recently been published by \citet{2016JBIS...69...278G}. It presents scenarios for antennas with sizes of 1--15\,km on both sides, transmitting MW power at 32\,GHz, achieving a data rate of Gbits/s ($10^9$\,bits/s). The antenna weight is mentioned as $40,000$\,kg, and the total space-ship weight is $10^7$\,kg. Clearly, if such masses and power can be sent to other stars, the question of communication will be trivial in comparison.

\subsection{\texttt{PyCom} software package}
We provide the Python-based software package \texttt{PyCom} as open source under the free MIT license\footnote{\hyperref[]{http://github.com/hippke/communication/}}. The repository provides function calls for the equations in this paper, a tutorial, and scripts to reproduce all figures.

\section{Conclusion}
In this work (paper I of the series), we have set the framework of data transfer between telescopes, using the example of a light-weight, power-limited probe at $\alpha$\,Cen. We have explored limiting factors such as extinction, noise, and technological constraints. We have calculated optimal frequencies and achievable data rates.

The Holevo bound gives an upper limit of a few bits per photon for realistic signal and noise levels from a communication between a meter-sized probe at $\alpha$\,Cen and a large (39\,m) telescope on earth. The achievable data rate is of order bits per second per Watt. For a probe with a size of a few meters, and photovoltaic energy of KW\,m$^{-2}$, power levels might be KW, resulting in data rates of order kbits/s. The optimal wavelength for a communication with $\alpha$\,Cen, at current technological levels, is 300\,nm (space-based receiver) to 400\,nm (earth-based) and increases with distance, due to extinction, to a maximum of $\approx3$\,$\mu$m to the center of the galaxy at 8\,kpc.

A critical requirement in this scheme is the coronagraphic suppression of the stellar background at the level of $10^{-9}$ within a few tenths of an arcsecond of the bright star, which has not been demonstrated yet. Further research on this topic is encouraged.

In paper II, the use of a stellar gravitational lens will be discussed. In paper III, we will relax technological constraints to explore the ultimate, most efficient interstellar communication scheme which yields insight into communication of more advanced life in the universe, if it exists.

\acknowledgments
The author is thankful to Ren\'{e} Heller, Duncan Forgan, John Learned and Tony Zee for helpful discussions, and to the Breakthrough Initiatives for an invitation to the Breakthrough Discuss 2017 conference at Stanford University.

\bibliographystyle{yahapj}

\end{document}